\documentclass[11pt]{article}
\usepackage{amssymb}
\usepackage{amsmath}
\usepackage[dvips]{epsfig}

\def\##1{\underline #1}
\def\=#1{\underline{\underline #1}}

\def\eps{\epsilon}
\def\veps{\varepsilon}

\def\vepso{\varepsilon_0}
\def\muo{\mu_0}
\def\ko{k_0}
\def\lambdao{\lambda_0}

\def\.{\mbox{ \tiny{$^\bullet$} }}

\def\ux{\#{u}_x}
\def\uy{\#{u}_y}
\def\uz{\#{u}_z}

\def\le{\left(}
\def\ri{\right)}
\def\les{\left[}
\def\ris{\right]}

\def\c#1{\cite{#1}}
\def\l#1{\label{#1}}
\def\r#1{(\ref{#1})}

\def\eps{\epsilon}

\def\muo{\mu_0}
\def\ko{k_0}
\def\lambdao{\lambda_0}

\def\.{\mbox{ \tiny{$^\bullet$} }}

\def\alph'{\alpha^{\prime}}

\def\ux{\#{u}_x}
\def\uy{\#{u}_y}
\def\uz{\#{u}_z}

\def\le{\left(}
\def\ri{\right)}
\def\les{\left[}
\def\ris{\right]}

\def\c#1{\cite{#1}}
\def\l#1{\label{#1}}
\def\r#1{(\ref{#1})}

\def\ux{\#u_x}
\def\uy{\#u_y}
\def\uz{\#u_z}

\def\aL{a_{\rm L}}
\def\aR{a_{\rm R}}
\def\rL{r_{\rm L}}
\def\rR{r_{\rm R}}
\def\tL{t_{\rm L}}
\def\tR{t_{\rm R}}

\def\psinc{\psi_{\rm inc}}
\def\cthe{\theta_{\rm inc}}

\def\ux{\#u_x}
\def\uy{\#u_y}
\def\uz{\#u_z}
\def\sp{\#s}
\def\pinc{\#p_+}
\def\pref{\#p_-}

\def\Ei{\#E_{\rm inc}(\#r)}

\def\Er{\#E_{\rm ref}(\#r)}

\def\Et{\#E_{\rm tr}(\#r)}

\def\M0{[\underline{\underline{M}}^\prime(0,\kappa,\psinc)]}

\begin{document}

\noindent {\bf On Calibration of a Nominal Structure--Property Relationship Model
for \\ Chiral  Sculptured Thin Films by Axial Transmittance Measurements }

\bigskip

\noindent J. A. Sherwin$^1$, A. Lakhtakia$^1$ and I. J. Hodgkinson$^2$

\bigskip

\noindent $^1$CATMAS~---~Computational and Theoretical Materials Science
Group\\ Department of Engineering Science and Mechanics\\
Pennsylvania State University, University Park, PA 16802--6812
\bigskip

\noindent $^2$University of Otago, Department of Physics \\ PO Box
56, Dunedin, New Zealand

\bigskip

\noindent ABSTRACT--A chiral sculptured thin film is fabricated
from patinal titanium oxide using the serial bideposition
technique. Axial transmittance spectrums are measured over a spectral
region encompassing the Bragg regime for axial excitation.
The same spectrums are calculated using a nominal
structure--property relationship model and the parameter space of
the model is explored for best fits of the calculated and measured
transmittances. Ambiguity arising on
calibrating the model against axial transmittance measurements
is shown to be resolvable using non--axial transmittance
measurements.


\vskip 1.0cm

\section{INTRODUCTION}\label{S:intro}

Among the nano--engineered materials recently identified by the US
National Research Council in a 1999 survey entitled {\em
Condensed--Matter and Materials Physics, Basic Research for
Tomorrow's Technology\/} as significant for scientific and
technological progress in the following two decades are sculptured
thin films (STFs) \c{LMBR96,Lmsec02}. These nano\-struc\-tur\-ed
inorganic materials with unidirectionally varying properties can
be designed and realized in a controllable manner using physical
vapor deposition \c{MVS00}--\c{ST01}. The ability to virtually
instantaneously change the growth direction of their columnar
morphology, through simple variations in the direction of the
incident vapor flux, leads to a wide variety  of microscopic
columns with two--or three--dimensional shapes.

At visible and infrared wavelengths, a single--section STF is a unidirectionally
nonhomogeneous
continuum with direction--dependent properties \c{VLbook}.
Several sections can be grown
consecutively into a
multisection STF \c{LMBR96,ST01}.
Chiral STFs
display the circular Bragg phenomenon in accordance with
their periodic nonhomogeneity along the thickness direction.
This phenomenon has been exploited to design, fabricate and test:
circular polarization filters and laser mirrors,
polarization--discriminatory handedness--inverters, and
spectral hole filters \c{Lmsec02,HWam01}.
Furthermore,
the porosity of STFs makes them attractive
for fluid concentration sensing applications,
as their optical response properties change in accordance with
the number density of infiltrant molecules, which has
been demonstrated theoretically as well as
experimentally with chiral STFs \c{LMSWH}. Other
optical, electronic, thermal, and biophysical applications
are also under investigation by many researchers \c{Lmsec02,HWam01},
and the future of STFs continues to
appear bright.

As STF technology matures from the proof--of--concept stage
towards the  marketable--devices stage,
experimentalists as well as theorists are increasingly challenged
to control and optimize the morphological and other
characteristics of chiral STFs for economically
attractive applications. On the
one hand, the recently developed serial bideposition (SBD)
technique \c{HWKLR}  is ideally suited for this purpose. It yields
chiral STFs with controlled morphology to enhance the so--called
local linear birefringence and optical activity. On the other
hand, a nominal model for structure--property relationships of
chiral STFs has been devised \c{SL2000,SL2001} and qualitatively
tested \c{LMSWH}.

In this communication, we report our first attempt at the
calibration of the nominal microscopic--to--macroscopic model
against transmittance spectrums measured when a chiral STF is
excited by a plane wave along its direction of nonhomogeneity. The
plan of this paper is as follows: The nominal model of a chiral
STF as an ensemble of ellipsoids is briefly outlined in section 2.
Model calibration involving the theoretical calculation of axial
transmittances and the experimental setup for their measurement
are outlined in section 3. Section 4 comprises the determination
of model parameters, a discussion of the ambiguity when
calibrating against axial transmittance measurements, and the
potential resolution of the ambiguity by non--axial transmittance
measurements.  An $\exp (-i \omega t)$ time--dependence is assumed
henceforth. All vectors are underlined once and all dyadics are
underlined twice.

\section{NOMINAL MODEL}
As any STF comprises clusters that are electrically small at optical
and lower frequencies,  it can be
considered as a material continuum at
those frequencies \c{LMBR96, VLbook}. Furthermore,
 STFs have a locally columnar morphology with
many void regions \c{MVS00}--\c{ST01}~---~which leads naturally to the
concept of local homogenization in order
to construct continuum models. In our model,
the deposited material as well as
the void regions are nominally conceived as confocal parallel
 ellipsoids in any
plane parallel to the substrate. The
Bruggeman formalism   is then used to estimate a reference permittivity
dyadic in
terms of two shape factors of
the ellipsoids, the bulk constitutive properties
of the deposited material, and the porosity. Infiltration
of the void regions
by some material can be handled by this model as also
the frequency--dependence of the
constitutive properties of the deposited materials. Calibration
against experimental data is an essential feature of this model.
As it has been discussed in detail elsewhere~\c{SL2001},
we give only a brief outline  here.

\subsection{Constitutive Relations}\label{S:consti}
The frequency--domain constitutive relations of a chiral STF
(after homogenization) are given by
\begin{align}\label{E:constit}
\underline{D}\left(\underline{r} \right) &= \vepso \,
\underline{\underline{\varepsilon}}\left(z \right) \.
\underline{E}\left(\underline{r} \right), \\
\underline{B}\left(\underline{r} \right) &= \muo \,\underline{H}
\left(\underline{r}\right),
\end{align}
where $\vepso$ and $\muo$ are the permittivity and permeability of
free space (i.e., vacuum). The nonhomogeneous relative
permittivity dyadic $\=\veps$ is written in terms of a homogeneous
permittivity dyadic $\=\varepsilon_{ref}^{o}$ and two rotation
dyadics as
\begin{equation}
\=\varepsilon \le z \ri= \=S_{z}\le z \ri \. \=S_{y} \le \chi\ri
\. \={\varepsilon}_{ref}^{o} \. \={S}_{y}^{-1}\le \chi\ri
\.\={S}_{z}^{-1}\le z \ri.
\end{equation}
Here,
\begin{eqnarray}\label{E:eref}
\=\veps_{ref}^{o} & = & \veps_{a} \#{u}_{z} \#{u}_{z} + \veps_{b}
\#{u}_{x} \#{u}_{x} + \veps_{c} \#{u}_{y} \#{u}_{y}, \\ & = &
n_{1}^{2}\#{u}_{x} \#{u}_{x} + n_{2}^{2} \#{u}_{z} \#{u}_{z} +
n_{3}^{2}\#{u}_{y} \#{u}_{y}, \notag \\
\underline{\underline{S}}_{y} \left( \chi \right) & = &
\underline{u}_{y} \underline{u}_{y} +
\left(\underline{u}_{x}\underline{u}_{x}+
\underline{u}_{z}\underline{u}_{z}\right) \cos \chi
+\left(\underline{u}_{z}\underline{u}_{x}-
\underline{u}_{x}\underline{u}_{z}\right) \sin \chi, \\
\label{E:eref3} \underline{\underline{S}}_{z} \left( z \right) & =
& \underline{u}_{z}\underline{u}_{z} +
\left(\underline{u}_{x}\underline{u}_{x}+
\underline{u}_{y}\underline{u}_{y} \right) \cos \le \pi z / \Omega
\ri + h \le
\underline{u}_{y}\underline{u}_{x}-\underline{u}_{x}\underline{u}_{y}\ri
\sin \le \pi z / \Omega \ri,
\end{eqnarray}
$2\Omega$ is the structural period, the angle of rise $\chi$
describes the elevation of the helicoidal columns above the $xy$
plane, whilst $n_{1,2,3}$ are the principal indexes of
refraction \c{HWH}. The structural handedness parameter
$h=1$ for right-- and $h=-1$ for left--handed
STFs. As the period is fixed {\em a priori\/} by the
deposition conditions, $\=\veps(z)$ can be
completely delineated,
provided that the reference permittivity dyadic
\begin{equation}
\=\veps_{ref} = \=S_{y} \le \chi \ri \. \=\veps_{ref}^{o} \.
\=S_{y}^{-1} \le \chi \ri
\end{equation}
 is known.

\subsection{Local Homogenization}\label{S:Homog}

Consider a homogenized composite medium (HCM)
 whose relative permittivity dyadic
$\=\veps_{HCM}$ is identical to that of the chosen chiral STF in the
limit $\Omega \to \infty$, i.e., $\=\veps_{HCM}=
   \={\veps}_{ref}$. The longest principal axes of all ellipsoids
in this HCM are aligned parallel to the unit vector
$\#{u}^{\prime}_{x} = \={S}_{y} \le \chi \ri \. \#{u}_{x}$, while
the smaller of the two remaining principal axes is aligned
parallel to the unit vector $\#{u}^{\prime}_{z} = \={S}_{y} \le
\chi \ri \. \#{u}_{z}$. The deposited material is isotropic
in bulk with relative permittivity scalar
$\varepsilon_{s}$~---~which can be
considered in our model as frequency--dependent~---~while the relative
permittivity of the void regions $\veps_{v}$ equals unity, of course.
    With respect to its
centroid, the surface of an ellipsoid may be described in
cartesian coordinates by the relation
\begin{equation}\label{E:one}
    {z^{\prime}}^{2} + \left( \frac{y^{\prime}}{\gamma_{2}} \right)^{2} + \left(
\frac{x^{\prime}}{\gamma_{3}} \right)^{2} = \delta^{2},
\end{equation}
where $\delta$ is a linear measure of the absolute size, while the
transverse aspect ratio $\gamma_{2}>1$ and the slenderness ratio
$\gamma_{3}>>1$ relate the three principal axes.

\indent The Bruggeman formalism involves the solution of the
dyadic equation~\cite{WLM97}
\begin{equation}\label{E:estimate}
f \,\=a_s+ (1-f) \,\,
\=a_v=\underline{\underline{0}} \, ,
\end{equation}
where $f$,\,($0 \leq f \leq 1$), is the volume fraction of the
film occupied by the deposited material and $\=0$ is the null dyadic. The
polarizability dyadics
$\=a_{s,v}$ are explicit
 functions of $\veps_{s,v}$, $\chi$, and
$\gamma_{2,3}$, and implicitly depend on $\=\veps_{ref}$ as well.
Standard iterative
methods detailed elsewhere~\cite{SL2001} are used to compute
$\=\veps_{ref}$.

\section{CALIBRATION OF THE NOMINAL MODEL}
Our first attempt to calibrate the described model involves the
excitation of a chiral STF by a normally incident plane
wave, and the measurement of the consequent transmittances.

\subsection{Axial Excitation}\label{S:five}

Let the region $0 \leq z \leq L$ be occupied by a chiral STF while the
regions $z \geq L$ and $z \leq 0$ are vacuous. An arbitrarily
polarized plane wave, with wavenumber $\ko=\omega \sqrt{\vepso
\muo}$ and wavelength $\lambda_0 =2\pi/\ko$,
 is normally incident on the  chiral STF from the lower
half--space $z \leq 0$.  This results in a plane wave reflected back
into the lower half--space, and a plane wave transmitted into the
upper half--space. The  electric field phasor in the lower half--space is given by
\begin{align}\label{E:phas1}
\underline{E}\left(\underline{r}\right) &=\left(a_{L}
\, \underline{u} _{+}+a_{R} \, \underline{u}_{-}\right) e^{i \ko z
}+ \left(r_{L} \, \underline{u} _{-}+r_{R}
   \, \underline{u}_{+}\right)
e^{-i\ko z} ;\quad z \leq 0 \, ,
\end{align}
where $\underline{u}_{\pm}=\left(\underline{u}_{x} \pm
i\underline{u}_{y} \right)/\sqrt{2}$.
   The transmitted electric field has the
phasor representation
\begin{align}\label{E:transm}
\underline{E}_{tr} \left( \underline{r} \right) &=
\left(t_{L} \, \underline{u}_{+} + t_{R} \, \underline{u}_{-}
\right) e^{i \ko \le z-L \ri};\quad z \geq L \, .
\end{align}
The quantities $a_{L}$ and $a_{R}$ are the amplitudes of the
left and right circularly polarized (LCP and RCP) components of the
incident plane
wave, while $r_{L,R}$ and $t_{L,R}$ are the analogous
amplitudes of the reflected and transmitted planewave components.

 A boundary value problem can be solved for the reflection and
transmission amplitudes in terms of the incidence wave amplitudes,
as discussed elsewhere in detail~\cite{VLbook}. For our present
purposes, the results are
compactly written in matrix form as
\begin{equation}
 \left[
\begin{matrix}
t_L\\ t_R
\end{matrix} \right]
= \left[
\begin{matrix}
t_{LL} & t_{LR} \\ t_{RL} & t_{RR}
\end{matrix} \right]
\left[
\begin{matrix}
a_L\\ a_R
\end{matrix}\right].
\label{deft}
\end{equation}
Of the 4 coefficients appearing in the foregoing $2 \times 2$
matrix, those with both subscripts identical refer to
co-polarized, while those with two different subscripts refer to
cross-polarized,  transmission.

\subsection{Experimental setup}
Chiral STFs were grown by SBD using Patinal titanium oxide S
granules. The details of SBD are discussed
elsewhere~\c{HWam01,WHL00}. It suffices to note here
that vapor is
incident from a single source under free streaming conditions at
the deposition angle $\chi_{v}$ (with respect to
the substrate). The deposition angle can assume
any value in the range $0^{\circ} < \chi_{v} \leq 90^{\circ}$,
but is typically set between $20^{\circ}$ and $30^{\circ}$.
The resultant morphology of the chiral STFs grown by
SBD has been described~\cite{HWKLR} as twisted
columns running normal to the substrate.

The essential features of the apparatus used to measure the
transmittance spectrums of an axially excited chiral STF have been
amply described by Wu et al.~\c{WHL00}. Most importantly, an
axially excited chiral STF displays the so--called circular Bragg
phenomenon. When $\lambda_0$ lies in the Bragg regime, incident
LCP light is preferentially reflected and incident RCP light is
preferentially transmitted by a structurally left--handed chiral
STF. All four transmittances $T_{LR} =\vert t_{LR}\vert^2$, etc.,
were measured as functions of $\lambda_0\in [400,800]$ nm at 2~nm
intervals. This range amply covered the Bragg regime.

\section{RESULTS OF CALIBRATION}
\subsection{Determination of $\veps_s$}
The bulk properties of a complex material such as Patinal titanium
oxide depend significantly on the conditions of preparation.
Hodgkinson et al.~\cite{HoWuCo2001} provided a new procedure to
measure $\veps_s$ by (i) growing a columnar thin film of Patinal
titanium oxide, (ii) measuring $\=\veps_{HCM}$ at
$\lambda_0=633$~nm, and (iii) inverting the Bragg--Pippard
formalism \c{BP1,BP2}. We modified that procedure by replacing the
Bragg--Pippard formalism by the Bruggeman formalism.

The principal refractive index $n_{1} \in \les 1.97,\, 2.03\ris$
when the deposition angle $\chi_{v}=20^{\circ}$ and
$\lambda_0=633$~nm \c{HWH}. This corresponds to $\veps_{b} \in\les
3.80,\, 4.12\ris$. The quantity $\veps_{s}$ was varied while
holding $\gamma_{3}=20$ fixed~---~in order to simulate columnar
morphology~---~and the Bruggeman formalism was
repeatedly implemented until
$\veps_b$ matched the estimated values of $n_{1}^2$. Agreement was
found for $\veps_{s} \in \les 5.8,\,6.15\ris$ at
$\lambda_0=633$~nm.

The center--wavelength of the Bragg regime of the specific chiral
STF was taken from the axial transmittance spectrums to be
$612$~nm. We linearized the functional relationship~\c{HoWuCo2001}
\begin{equation}
\veps_s(\lambdao) =\veps_s\Bigg\vert_{\lambdao=633~{\rm nm}} \,
 \left[ 1 + D \le \frac{1}{\lambdao^{2}}
-\frac{1}{633^{2}} \ri \right]^2
\end{equation}
to take the form
\begin{equation}
\veps_s(\lambdao) =\veps_s\Bigg\vert_{\lambdao=633~{\rm nm}}\,
\left[ 1 - \frac{2 D}{633^{3}}\le \lambdao - 633 \ri \right]^2,
\end{equation}
in order to estimate $\veps_s$ at $\lambdao=612$~nm. Based on
recent measurements of the refractive index of titanium oxide
films~\c{emlink}, we determined $D=5.0 \times 10^{4}$~nm$^{-2}$.
Absorption was not considered in the described scheme, and so the
determination of $\veps_s$ had to be augmented by the addition of
a suitable imaginary part.

\subsection{Axial Transmittances}
A structurally left--handed chiral STF of thickness $L=5200$~nm
with half--period $\Omega =
173$~nm was fabricated keeping $\chi_{v}=20^{\circ}$ fixed. The
film was deposited on a substrate about 1~mm thick with refractive
index $n_{sub} = 1.52$. The axial transmittances $T_{LL}$,
$T_{RR}$, $T_{LR}$, and $T_{RL}$ were measured over the range
$\lambdao\in \les 400,\,800\ris$~nm.

The parameters $f$, $\chi$ and $\gamma_{2}$ were varied so that
the calculated transmittances $T_{RR}$, $T_{LL}$, $T_{RL}$ and
$T_{LR}$ would best fit the measured data while $\gamma_{3}$ was
held fixed at 20. This value of $\gamma_{3}$ was chosen because we
wanted to simulate the locally columnar morphology.  As
no data was available concerning \rm{I}m$[\veps_{s}]$,
the value $0.012$ was chosen for this quantity so that the
calculated spectral averages
of $T_{LL}$ approximately matched their
 measured
counterparts  in both vicinities of the Bragg regime.

Sample results are presented in Figure 1. Good agreement between
measured and calculated co--polarized transmittance spectrums is found over the
Bragg regime which is approximately 40~nm wide.
 We also note that the predicted spectrums of
$T_{LR}$ and $T_{RL}$ are the same   \c{VLbook,McC},
but the measured spectrums differ. The experimentally
observed difference between $T_{LR}$ and $T_{RL}$
is because the refractive index of the substrate is not the same as that
of a lid covering the other face of the film.
Anyhow, both cross--polarized transmittances
are negligible in comparison to $T_{RR}$, and can be ignored
therefore.

That  portion of the $\gamma_2$--$\chi$ space where
matches between predicted and measured axial transmittances occur
is presented in Figure 2. Projections of the solution regions of
the $\gamma_2$--$\chi$--$f$ space onto the $\gamma_2$--$f$ and the
$\chi$--$f$ planes are not shown, because we found that  the value of $f$ which creates a match at the
center--wavelength of the Bragg regime is nearly fixed. The
center--wavelength shifts with $f$  for the
measured co--polarized transmittances to be
adequately matched.

The center--wavelength  of the
Bragg regime also shifts very
slightly with variations in
$\chi$, but the bandwidth of the Bragg regime depends
more strongly on $\chi$. From numerous simulation
trials, it appears that
quite specific values of $f$ and $\gamma_2$ are required to match
both the center--wavelength and the bandwidth of the Bragg regime,
when $\chi$ is fixed.

Our model predicts two disjoint regions $\gamma_2$--$\chi$ space
where good agreement between the model and the measurements is found.
This leads, for example, to multiple values of $\chi$
corresponding to a specific value of $\gamma_{2}$. The volume
fraction was found to lie in the fairly narrow range $0.51 \leq f
\leq 0.63$~---~the lowest value corresponding to $\chi =
25^{\circ}$ in Figure 2a (Region 1), and the highest value
corresponding to $\chi = 90^\circ$ in Figure 2b (Region 2).

\subsection{Resolution of ambiguity}
Axial transmittance data can assist in the calibration of our
model, but it does leave the ambiguity between Regions 1 and 2
unresolved. The disparity between the values of $\chi$ in the two
Regions is quite large. Available scanning electron micrographs do
not give clear indication of $\chi$ due to shadowing effects.
Furthermore, the ellipsoidal model used here is nominal, so that
$\chi$ itself may only be loosely connected to the actual
microstructure. We therefore examined the theoretical responses of
chiral STFs to non--axial excitation by plane waves in order to
resolve the ambiguity.

Let the chiral STF be
excited by a plane wave propagating at an angle $\theta_{inc}$ to
the $z$ axis and at an angle $\psi_{inc}$ (in the $xy$ plane)
to the $x$ axis.  The electric field phasor associated with the
incident plane wave can be represented as~\c{VL2000}
\begin{eqnarray}
\nonumber \Ei&=& \les \frac{\le i\sp - \pinc \ri}{\sqrt{2}} \, \aL
- \frac{\le i\sp +\pinc \ri}{\sqrt{2}} \, \aR \ris \\ \l{incEphas}
& & \qquad \times \, e^{\les i\kappa \le x\cos\psinc + y\sin\psinc
\ri \ris} e^{i\ko z\cos\cthe} \, , \, \, z \leq 0 \,,
\end{eqnarray}
where
\begin{eqnarray}
&&\kappa =
\ko\sin\cthe\,,\\
&&\sp=-\ux\sin\psinc + \uy \cos\psinc \, ,\\
&&\#p_\pm = \mp\le \ux
\cos\psinc + \uy \sin\psinc \ri \cos\cthe + \uz \sin\cthe \, .
\end{eqnarray}
The
electric field phasor of the reflected plane wave can be
represented as
\begin{eqnarray} \nonumber \Er&=& \les \frac{-\le
i\sp - \pref \ri}{\sqrt{2}} \, \rL + \frac{\le i\sp + \pref
\ri}{\sqrt{2}} \, \rR \ris \\ \l{refEphas} & & \qquad \times \,
e^{\les i\kappa \le x\cos\psinc + y\sin\psinc \ri \ris} e^{-i\ko
z\cos\cthe} \, \, , z \leq 0 \,,
\end{eqnarray}
and that of the transmitted plane wave as
\begin{eqnarray}
\nonumber \Et&=& \les \frac{\le i\sp - \pinc \ri}{\sqrt{2}} \, \tL
- \frac{\le i\sp + \pinc \ri}{\sqrt{2}} \,  \tR \ris \\
\l{trEphas} & & \qquad \times \, e^{\les i\kappa \le x\cos\psinc +
y\sin\psinc \ri \ris} e^{i\ko \le z-L \ri \cos\cthe} \, , \, \, z
\geq L \, .
\end{eqnarray}

A boundary value problem similar to that for axial excitation was
solved to obtain the four coefficients of \r{deft}. The angles of
incidence were set at $\theta_{inc} = 60^{\circ}$ and $\psi_{inc}
= 0^{\circ}$, while $\veps_{s} = 5.95 + 0.012 i$ was chosen
independent of the wavelength. The spectrums of $T_{LL}$ are
presented in Figure 3 for the following three cases:
\begin{itemize}
\item[A.]
$\gamma_{2} = 1.06$, $\chi=45^\circ$, $f=0.592$ (Region 1),
\item[B.]
$\gamma_{2} = 2.70$, $\chi=45^\circ$, $f=0.585$ (Region 2),  and
\item[C.]
$\gamma_{2} = 1.47$, $\chi=90^\circ$, $f=0.630$ (Region 2).
\end{itemize}
These results clearly indicate that the non--axial transmittances
of chiral STFs with quite different microstructural parameters
will be different from one another even though their axial
transmittances are virtually indistinguishable. The Bragg
phenomenon virtually disappears for Case C, but not for Cases A
and B. The value of $\gamma_{2}$ can also be distinguished through
non--axial transmittance measurements, as is obvious from the
differences between Cases A and B in Figure 3.

\section{CONCLUDING REMARKS}
A chiral sculptured thin film was fabricated from Patinal titanium
oxide using the
 serial bideposition technique, and axial
transmittance spectrums were measured over a band of wavelengths
encompassing the Bragg regime for axial excitation. The same
transmittances were simulated using a nominal structure--property
relationship model. The $\gamma_{2}$--$\chi$--$f$
parameter space was explored for best fits of the calculated
transmittances to the measured values.

The following conclusions were arrived at:
\begin{itemize}
\item  Porosity of
chiral STFs was reaffirmed by our model.
\item  Axial transmittance data can not completely
resolve ambiguities in the calibration of the model.

\item Non--axial transmittance data appears
crucial to the resolution of the ambiguities.
\end{itemize}
We expect to present a detailed calibration
scheme in our future publications.

\newpage
\begin{center}
{\bf Figure Captions}
\end{center}

\noindent {\bf Figure 1.} Computed and measured spectrums of the axial
trasnmittances of a chiral STF: (a) $T_{LL}$, (b) $T_{RR}$,  (c)
$T_{LR}$, and (d) $T_{RL}$. Computations were carried out with
$\gamma_3=20$, $\gamma_2=1.06$, $\Omega=173$nm, $L=30\Omega$,
$f=0.579$, $\chi=47^{\circ}$, and $\veps_{s}=6.3 +0.012 i$.\\

\noindent {\bf Figure 2.} Regions 1 and 2 of the $\gamma_{2}-\chi$
space.  The lower and upper bounds are delineated by
$\veps_{s}=5.95+0.012i$ (broken line) and
$\veps_{s}=6.30+0.012i$ (solid line).\\

\noindent {\bf Figure 3.} Calculated spectrums of $T_{LL}$ for
non--axial excitation of a chiral STF. These correspond to Cases A
(dot--dashed), B (dashed), and C (solid) described in Section 4.3.
Computations were carried out with $\gamma_3=20$, $\eps_{s} = 5.95
+ 0.012 i$, $\Omega=173$~nm, and $L=30\Omega$.\\

\newpage
\begin{figure}\centering
\epsfig{file=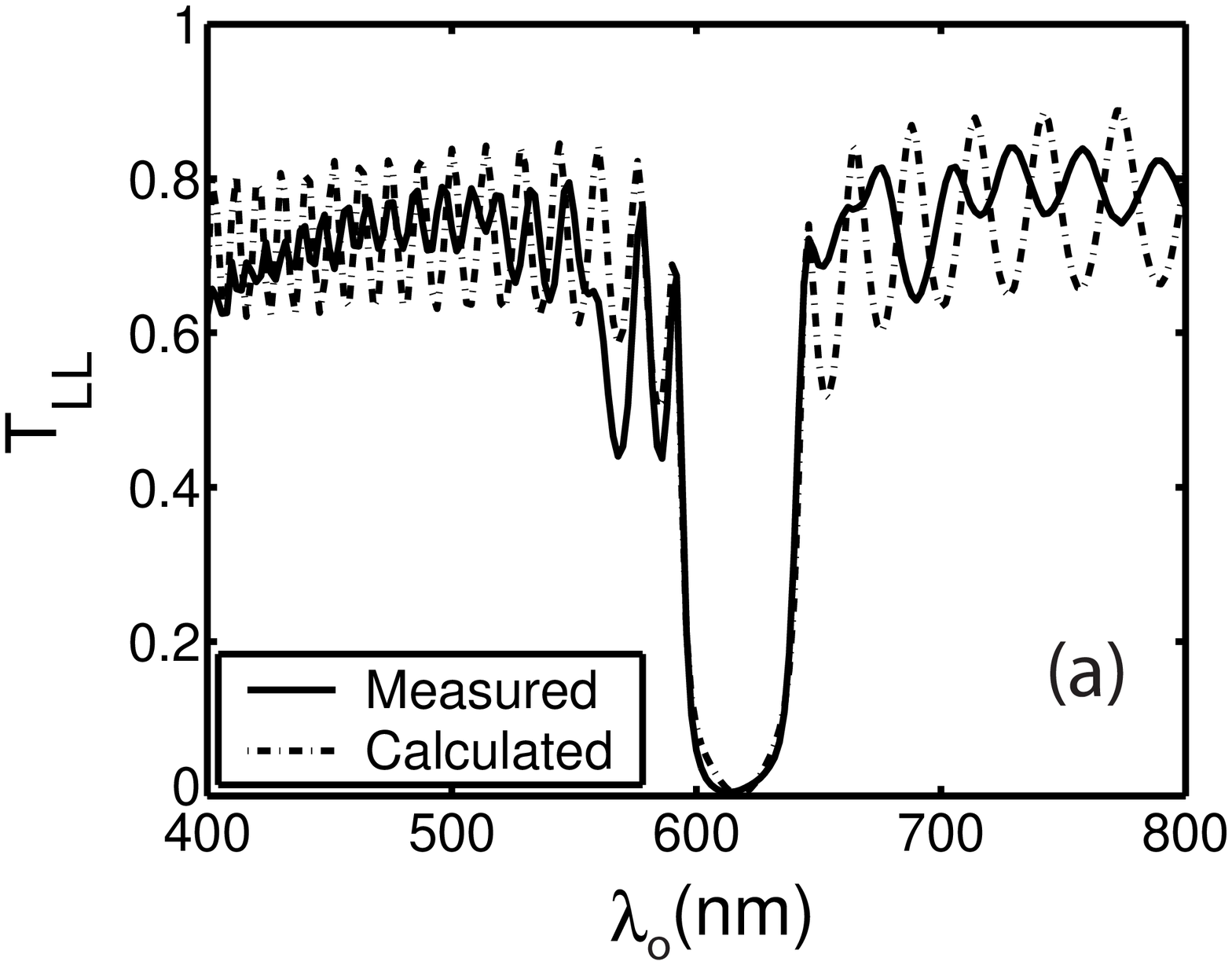, height=5.0cm,width=6.0cm,bbllx =
150, bblly = 73, bburx = 641, bbury =505} \hskip 1cm
\epsfig{file=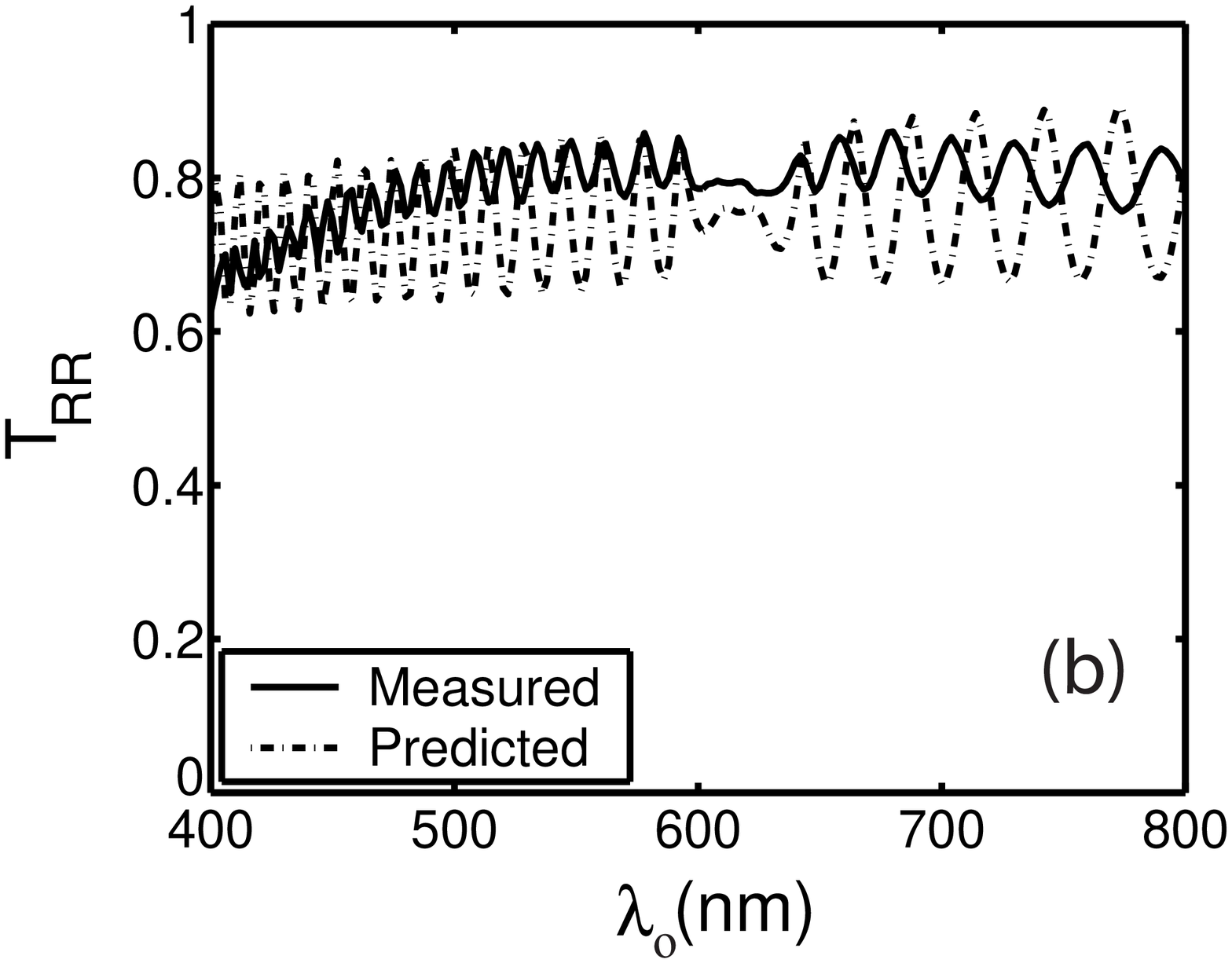, height=5.0cm,width=6.0cm,bbllx =
150, bblly = 73, bburx = 641, bbury =505} \vskip 1.0cm
\epsfig{file=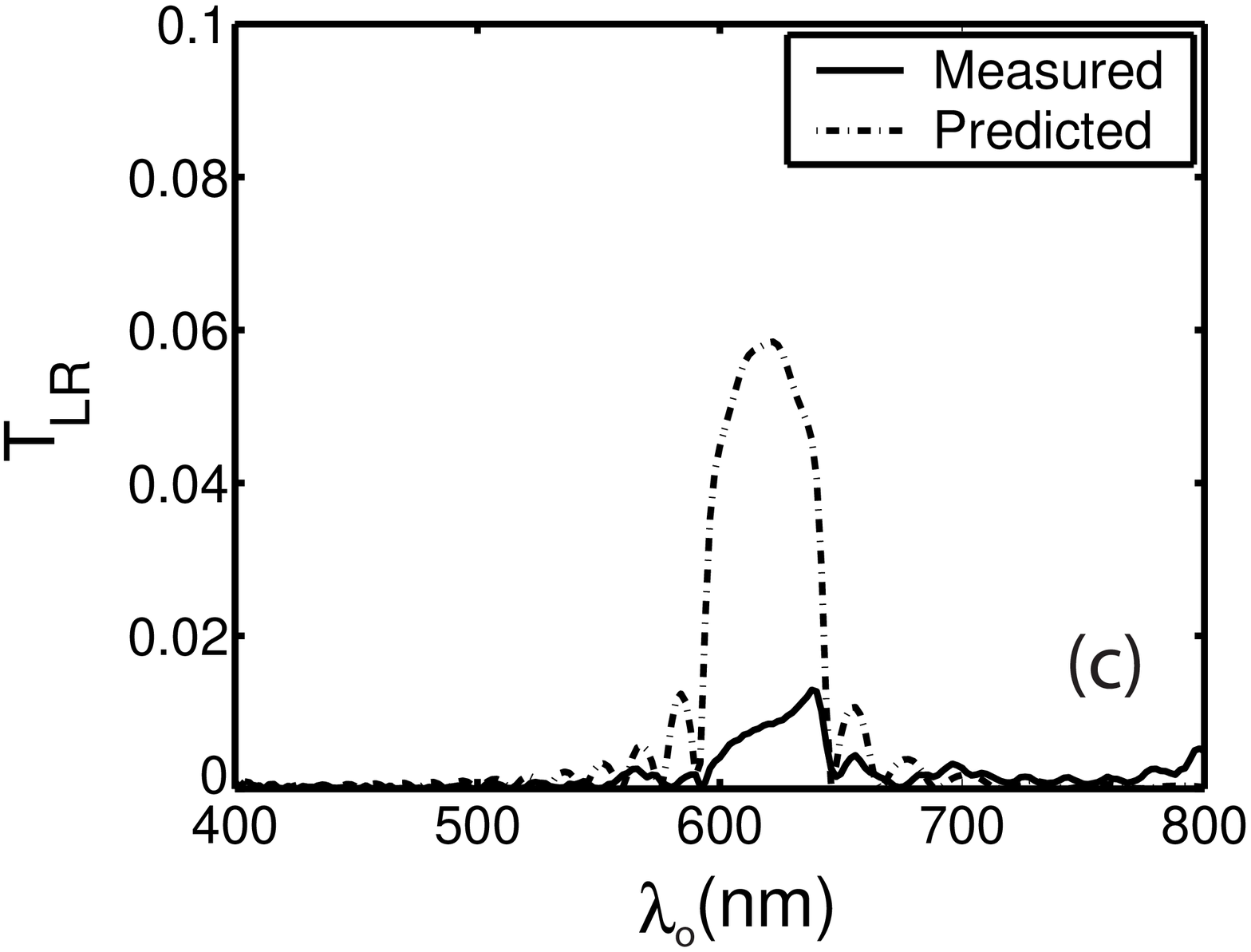, height=5.0cm,width=6.0cm, bbllx =
150, bblly = 73, bburx = 641, bbury = 505}
 \hskip 1cm
\epsfig{file=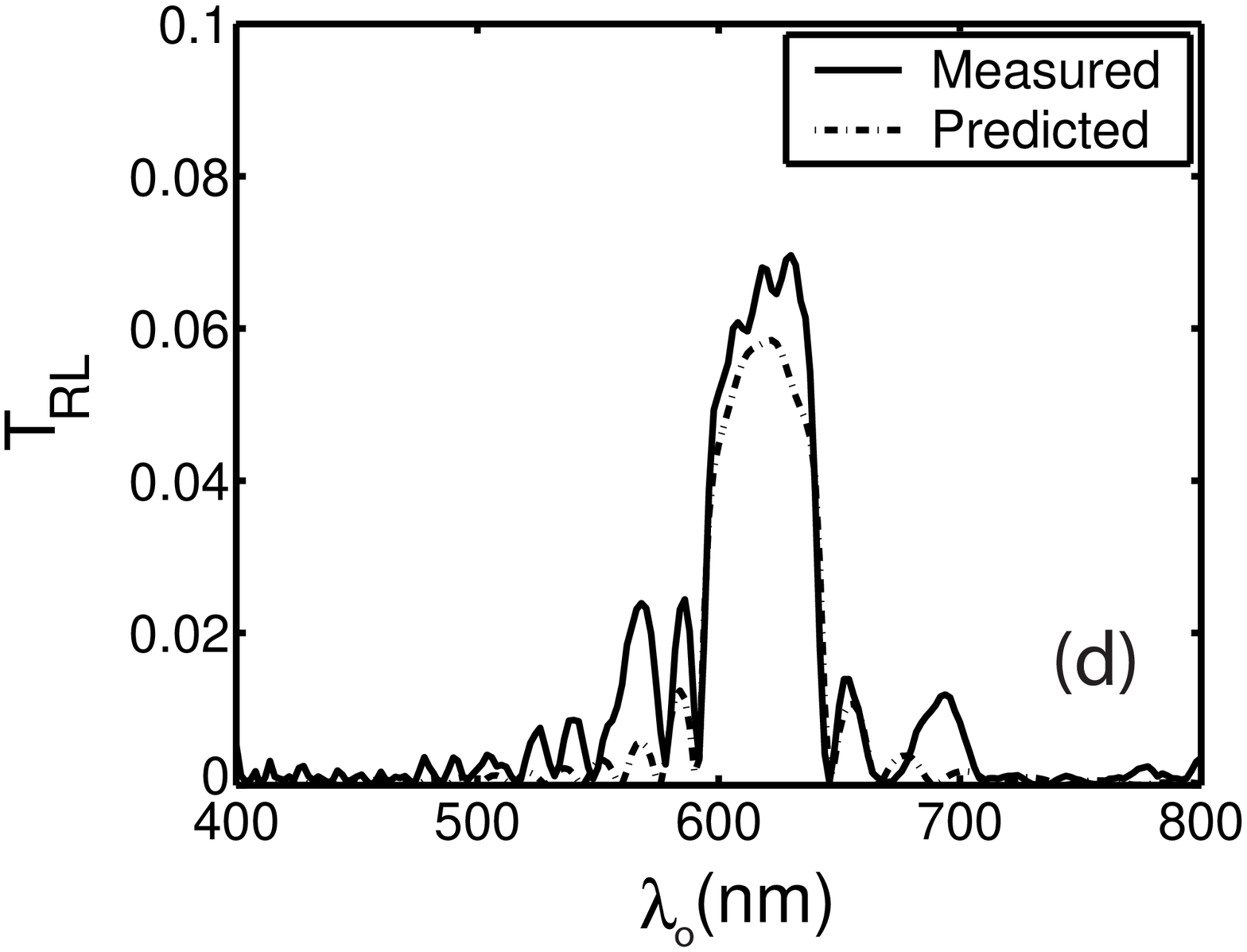, height=5.0cm,width=6.0cm,bbllx =
150, bblly = 73, bburx = 641, bbury =505} \vskip
5cm\caption{Computed and measured spectrums of the axial
trasnmittances of a chiral STF: (a) $T_{LL}$, (b) $T_{RR}$,  (c)
$T_{LR}$, and (d) $T_{RL}$. Computations were carried out with
$\gamma_3=20$, $\gamma_2=1.06$, $\Omega=173$nm, $L=30\Omega$,
$f=0.579$, $\chi=47^{\circ}$, and $\veps_{s}=6.3 +0.012 i$. }
\end{figure}

\newpage
\begin{figure}\centering
\vskip 1.0cm \epsfig{file=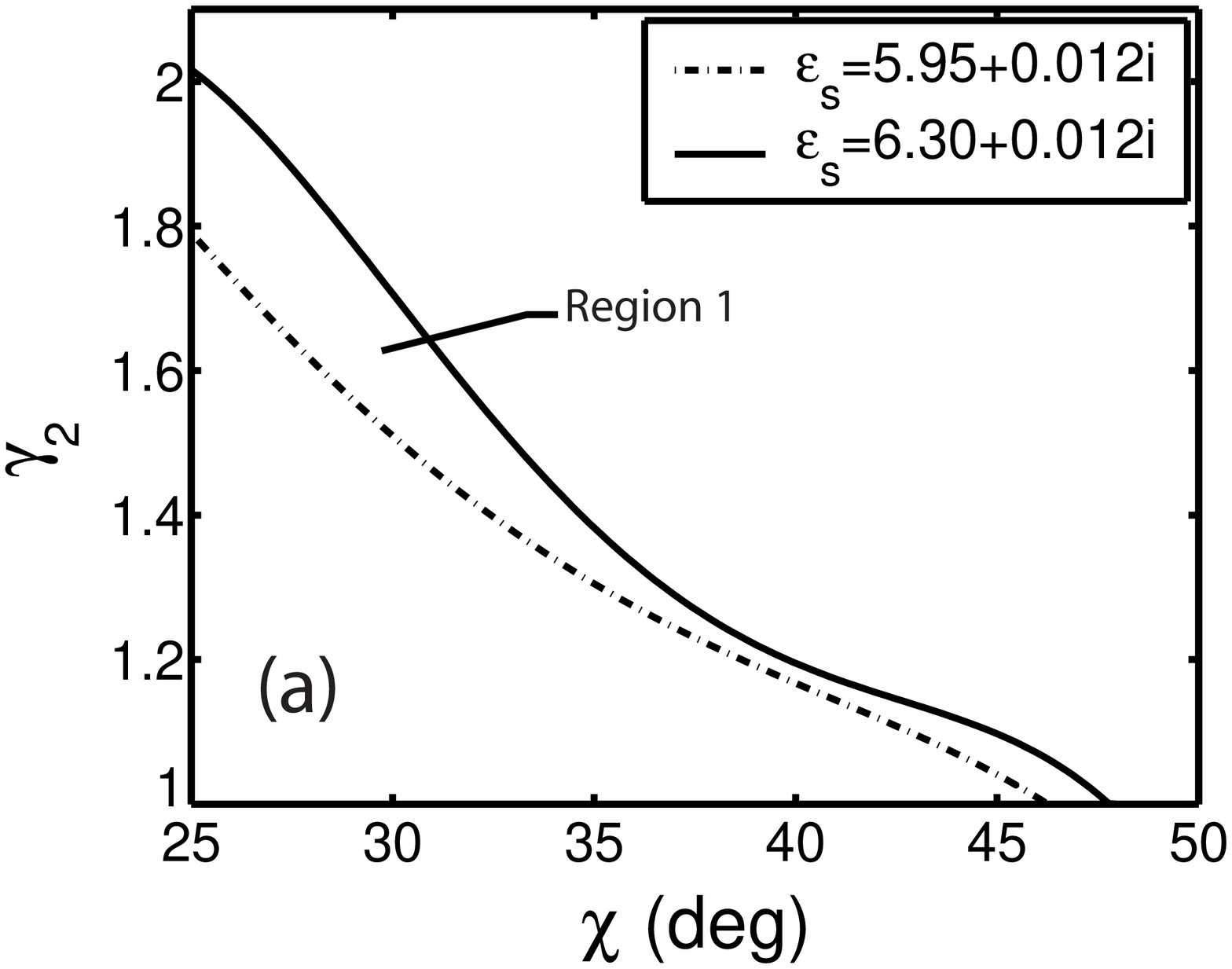,
height=5.0cm,width=6.0cm,bbllx = 150, bblly = 73, bburx = 641,
bbury =505} \vskip 1.0cm \epsfig{file=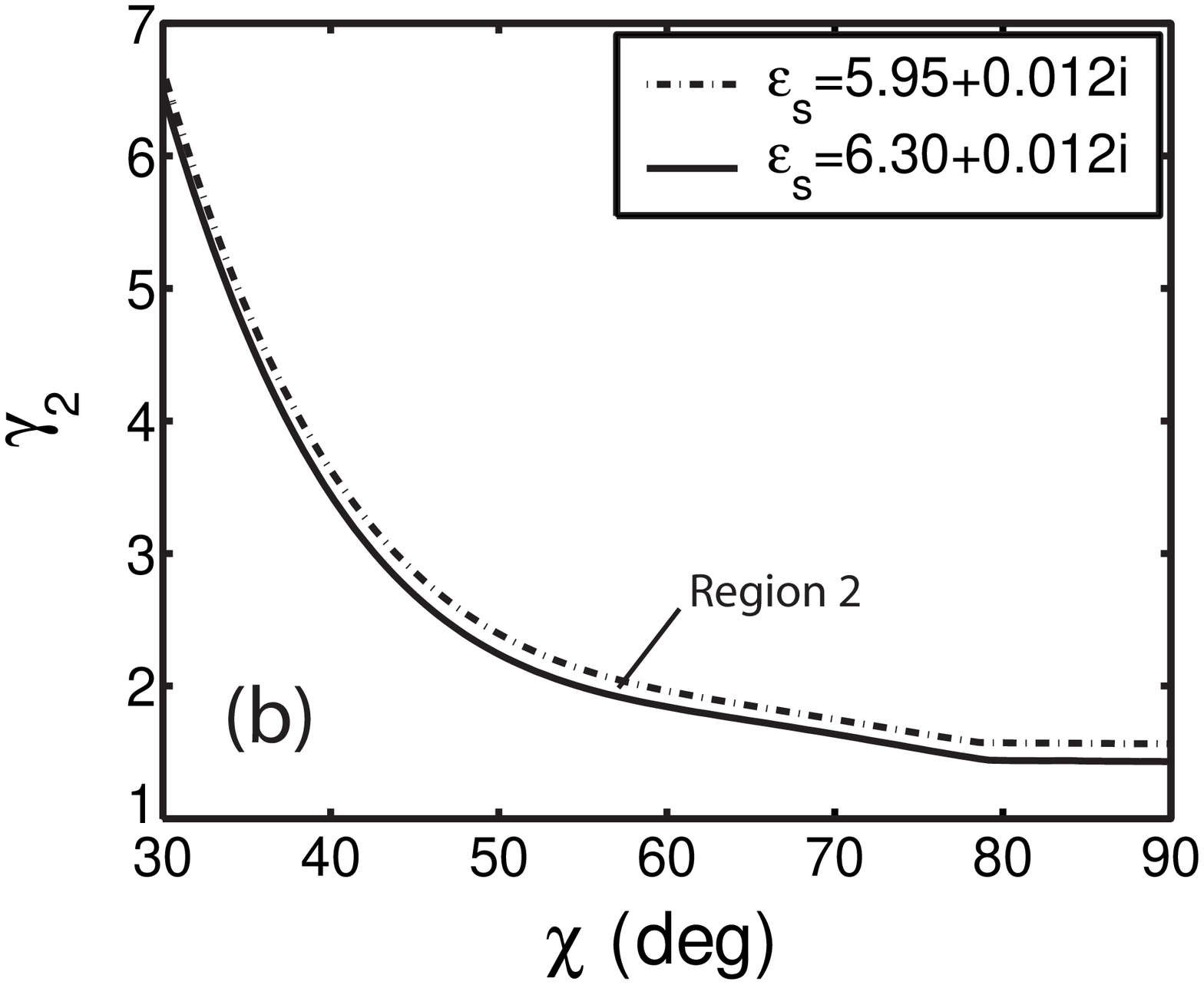,
height=5.0cm,width=6.0cm,bbllx = 150, bblly = 73, bburx = 641,
bbury =505} \vskip 5cm\caption{Regions 1 and 2 of the
$\gamma_{2}-\chi$ space.  The lower and upper bounds are
delineated by $\veps_{s}=5.95+0.012i$ (broken line) and
$\veps_{s}=6.30+0.012i$ (solid line).}
\end{figure}

\newpage
\begin{figure}\centering
\vskip 1.0cm \epsfig{file=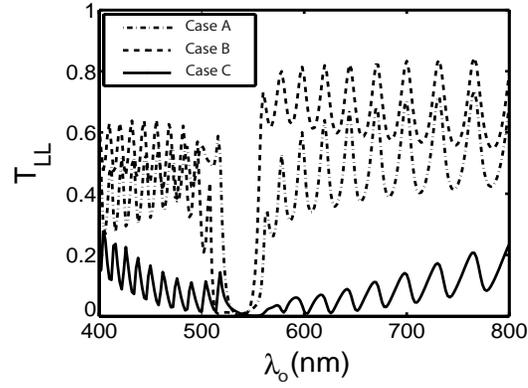,
height=5.0cm,width=6.0cm,bbllx = 150, bblly = 73, bburx = 641,
bbury =505} \vskip 5cm \caption{Calculated spectrums of $T_{LL}$
for non--axial excitation of a chiral STF. These correspond to
Cases A (dot--dashed), B (dashed), and C (solid) described in
Section 4.3. Computations were carried out with $\gamma_3=20$,
$\eps_{s} = 5.95 + 0.012 i$, $\Omega=173$~nm, and $L=30\Omega$.}
\end{figure}
\end{document}